\documentclass[twocolumn, aps, floatfix, superscriptaddress, showpacs]{revtex4-1}
\usepackage{amsmath, amssymb}
\usepackage{braket}
\usepackage{color}
\usepackage{graphicx}
\usepackage{dcolumn}
\begin{document}
\date{\today}
\title{Finite temperature thermodynamic properties of the spin-1 nematics in an applied magnetic field}
\author{Katsuhiro Tanaka}
\affiliation{Department of Basic Science, University of Tokyo, Meguro, Tokyo 153-8902, Japan}
\author{Chisa Hotta}
\affiliation{Department of Basic Science, University of Tokyo, Meguro, Tokyo 153-8902, Japan}
\begin{abstract}
We study numerically the thermodynamic properties of the spin nematic phases in a magnetic field in the spin-1 bilinear-biquadratic model. 
When the field is applied, the phase transition temperature once goes up and then decreases rapidly toward zero, 
which is detected by the peak-shift in the specific heat. 
The underlying mechanism of the reentrant behavior is the entropic effect. 
In a weak field the high temperature paramagnetic phase rapidly loses its entropy while the ferroquadrupolar nematic phase remains robust
by modifying the shape of the ferroquadrupolar moment. 
This feature serves as a fingerprint of generic ferroquadrupolar phases, while it is not observed for the case of antiferroquadrupoles.
\end{abstract}
\pacs{}
\maketitle
\textit{Introduction.}
Among symmetry broken phases in solids, 
those of higher order multipolar degrees of freedom are hard to study 
and have been often referred to as ``hidden orders''. 
This is because their order parameters are not linearly coupled to external fields or forces, 
and thus are extremely difficult to characterize by the conventional experimental probes, 
as well known in heavy fermion examples like $\mathrm{CeB_{6}}$~\cite{Effantin1985_JMagMagMater_47-48_145,Tayama1997_JPhysSocJpn_66_2268}, $\mathrm{URu_{2}Si_{2}}$~\cite{Buyers1996_PhysicaB_223-224_9,Shah2000_PhysRevB_61_564}. 
Still, making use of the coupling between wave function and the crystal lattice distortion, 
the the electronic quadrupolar orderings in $\mathrm{CeB_{6}}$ is detected by the elasticity measurements~\cite{Suzuki1998_JPhysSocJpn_67_4243}. 
The elasto-resistivity measurements are also performed for the clarification of the orbital nematics 
in iron pnictide~\cite{Kuo2013_PhysRevB_88_085113}. 
\par
In quantum magnets, the quadrupolar orderings of localized spins are often referred to as spin nematics. 
Unlike the true hidden orders, the quadrupolar spin moments are already well-defined in theories~\cite{Andreev1984_SovPhysJETP_60_267}. 
Nevertheless, the spin nematics are still ``hidden'' in the sense that 
they are often invisible to local magnetic probes like neutron scattering or magnetic resonances 
and only show featureless paramagnetic-like responses to static magnetic field. 
Efforts on measuring the dynamical quantities have been made~\cite{Lauchli2006_PhysRevLett_97_087205,Tsunetsugu2006_JPhysSocJpn_75_083701,Tsunetsugu2007_JPhysCondensMatter_19_145248,Sato2009_PhysRevB_79_060406,Podolsky2009_PhysRevB_79_140402,Sato2011_PhysRevB_83_064405,Sato2011_JPhysConfSer_320_012014,Shindou2013_PhysRevB_87_054429,Nawa2013_JPhysSocJpn_82_094709,Smerald2013_PhysRevB_88_184430,Starykh2014_PhysRevB_89_104407,Smerald2015_PhysRevB_91_174402,Furuya2017_PhysRevB_95_014416,Orlova2017_PhysRevLett_118_247201,Furuya2018_PhysRevB_97_104411,Ramos2018_PhysRevB_98_094431}, 
e,g, in the nuclear magnetic resonance~\cite{Tsunetsugu2007_JPhysCondensMatter_19_145248,Sato2009_PhysRevB_79_060406,Podolsky2009_PhysRevB_79_140402,Sato2011_PhysRevB_83_064405,Sato2011_JPhysConfSer_320_012014,Nawa2013_JPhysSocJpn_82_094709,Orlova2017_PhysRevLett_118_247201} 
and inelastic neutron scattering~\cite{Smerald2013_PhysRevB_88_184430,Smerald2015_PhysRevB_91_174402}, 
or the electron spin resonance~\cite{Furuya2017_PhysRevB_95_014416,Furuya2018_PhysRevB_97_104411}, 
while experiments and theories are practically difficult to reconcile. 
\par
Spin nematics can resort to Landau's approaches on second order phase transitions, 
where some anomalies are found in the magnetocaloric 
and in the ac magnetic susceptibility measurements~\cite{Schmidt2007_PhysRevB_76_125113,Kohama2019_PNAS_116_10686,Skoulatos2019_PhysRevB_100_014405}. 
In theories, finite temperature phase diagrams of spin-1 nematics at zero field are studied 
for the square lattice~\cite{Harada2001_JPhysSocJpn_70_13,Harada2002_PhysRevB_65_052403} and the triangular lattice~\cite{Kaul2012_PhysRevB_86_104411,Voll2015_PhysRevB_91_165128} by the quantum Monte Carlo (QMC) simulation, and for the triangular lattice by the variational method~\cite{Stoudenmire2009_PhysRevB_79_214436}. 
However, the basic information on how the quadrupolar moments of spin-1 respond to a magnetic field 
and they modify the finite temperature properties remain unexplored. 
\par
In this paper, we examine this issue by applying the Monte Carlo simulations 
to the spin-1 bilinear-biquadratic (BLBQ) model, a canonical model for spin nematics. 
A pure quantum quadrupole is represented by a director $\boldsymbol{d}$, a real vector pointing perpendicular 
to the flucuating spin moments, which does not couple to a magnetic field. 
However, the imaginary component of $\boldsymbol{d}$ relevant to the emergent dipole moment 
couples to the field, which modifies the shape of the quadrupole to those flutuating in-plane~\cite{Lauchli2006_PhysRevLett_97_087205}. 
Since this modification keeps the spin nematic state robust against the weak field, 
the transition temperature once slightly goes up and then starts to decrease at stronger field. 
The former behavior reminds us of the Pomeranchuk effect in $^3$He~\cite{Pomeranchuk1950_ZhEkspTeorFiz_20_919} 
whose liquid and solid phases may roughly correspond to our nematic and paramagnetic phases, respectively. 
In fact, the paramagnetic phase rapidly loses its entropy with an applied field and gives way to the nematic phase, 
just like decreasing the pressure in $^3$He will transform the solid to liquid by releasing the entropy. 
The reentrant effect is a general feature of the ferroquadrupolar nematic phase 
as it is observed also in a square lattice, whereas not in the antiferroquadrupolar phases. 
\par
\textit{Spin-1 bilinear-biquadratic model.}
We deal with the spin-1 BLBQ model on the triangular and square lattices in a magnetic field;
	\begin{align}
		\mathcal{H}
	= \sum_{\braket{i, j}} 
			\left[
			J \hat{\boldsymbol{S}}_{i} \cdot \hat{\boldsymbol{S}}_{j}
			+ K \left( \hat{\boldsymbol{S}}_{i} \cdot \hat{\boldsymbol{S}}_{j} \right)^{2}
			\right]
		- h \sum_{i = 1}^{N} \hat{S}_{i}^{z},
	\label{ham}
	\end{align}
where $\hat{\boldsymbol{S}}_{i}$ is the spin-1 operator on site-$i$ with $i=1$--$N$,
and $J$, $K$, and $h$ denote the Heisenberg (bilinear) and the biquadratic (BQ) interactions, 
the magnetic field applied parallel to the $z$-axis, respectively.
Hereafter, $J$ and $K$ are normalized as $J^{2} + K^{2} = 1$.
\par
Semiclassical approaches using the variational method revealed that 
in the absence of a magnetic field, 
the ground state of Eq.~(\ref{ham}) has a FQ order 
for $K/J < \tan^{-1}(-2)$ with $J > 0$, or $K < J\leq 0$ for the triangular lattice,
and $K < J < 0$ for the square lattice~\cite{Lauchli2006_PhysRevLett_97_087205,Penc2011_IFM_Chap13_SpinNematics,Toth2012_PhysRevB_85_140403}, 
which agree qualitatively well with those from the fully quantum approaches.
When a magnetic field is applied along the $z$-axis, 
the FQ state acquires a small but finite magntic moment along the $z$-axis 
while retains its quadrupolar moment in the $xy$-plane, 
but finally turns into a fully polarized magenetic phase at $h=z(J-K)$, where $z$ is the coordination number~\cite{Ivanov2003_PhysRevB_68_052401,Lauchli2006_PhysRevLett_97_087205,Penc2011_IFM_Chap13_SpinNematics,coordination_number}.
\par
\textit{Monte Carlo method with semiclassical SU(3) approximation.} 
We employ the semiclassical SU(3) approximation combined with classical Monte Carlo methods (sSU(3)-MC)~\cite{Stoudenmire2009_PhysRevB_79_214436}. 
The wave functions are approximated by the direct product form of the one-body wave functions as
	\begin{align}
		\ket{\Psi} = \bigotimes_{i = 1}^{N} \ket{\psi_{i}}, \
		\ket{\psi_{i}} = \sum_{\alpha = x, y, z} d_{i, \alpha} \ket{\alpha}_{i},
         \label{wf}
	\end{align}
where $d_{i, \alpha}$ is the complex coefficient satisfying $\left| \boldsymbol{d}_{i} \right| = 1$.
The time-reversal invariant basis states $\ket{\alpha}$ are given as
	\begin{align}
		\ket{x} = \dfrac{\mathrm{i}\left(\ket{+1}-\ket{-1}\right)}{\sqrt{2}}, \ 
		\ket{y} = \dfrac{\ket{+1} + \ket{-1}}{\sqrt{2}} , \ 
		\ket{z} = -\mathrm{i}\ket{0},
	\end{align}
where $\ket{n}$ ($n = 0, \pm 1$) is the spin-1 state with $S^{z} = n$.
The energy evaluated using this wave function, $E_{\text{sSU(3)}}=\braket{\Psi|\mathcal{H}|\Psi}$, is given as 
	\begin{align}
		\label{eq:hamiltonian_1body}
		E_{\text{sSU(3)}}
	& = \sum_{\braket{i, j}}
			\left[
				J \left| \boldsymbol{d}_{i}^{*} \cdot \boldsymbol{d}_{j} \right|^{2}
			+ \left( K - J \right)
					\left| \boldsymbol{d}_{i} \cdot \boldsymbol{d}_{j} \right|^{2}
			\right] \notag \\
	&\quad + \mathrm{i}h \sum_{i = 1}^{N} \left( \boldsymbol{d}_{i}^{*} \times \boldsymbol{d}_{i} \right)^{z}
		+ \text{const.}. 
	\end{align}
A set of parameters $\left\{ d_{i, \alpha} \right\}_{i = 1, \cdots, N}$
are updated by the standard classical MC sampling with the canonical ensemble of 
$\exp{\left( -\beta E_{\text{sSU(3)}} \right)}$, where $\beta=(k_{\mathrm{B}}T)^{-1}$ is the inverse temperature.
Spin moment whose $z$-component appears in the Zeeman term of Eq.~(\ref{eq:hamiltonian_1body}) is given explicitly as 
	\begin{align}
		\boldsymbol{S}_{i}
	  = \braket{\psi_{i}|\hat{\boldsymbol{S}}_{i}|\psi_{i}}
	& = - \mathrm{i}
				\begin{pmatrix}
					d_{i, y}^{*} d_{i, z} - d_{i, z}^{*} d_{i, y} \\
					d_{i, z}^{*} d_{i, x} - d_{i, x}^{*} d_{i, z} \\
					d_{i, x}^{*} d_{i, y} - d_{i, y}^{*} d_{i, x} \\
				\end{pmatrix}.
	\end{align}
Spin quadrupolar operator, 
$\hat{Q}_{i}^{\alpha\beta} = \hat{S}_{i}^{\alpha}\hat{S}_{i}^{\beta} + \hat{S}_{i}^{\beta}\hat{S}_{i}^{\alpha} - 2S(S + 1)/3\delta_{\alpha\beta}$, 
is a rank-2 traceless symmetric tensor, 
and its vector representation for five linearly independent components, 
$\hat{\boldsymbol{Q}}_{i}= (\hat{Q}_{i}^{x^{2}-y^{2}}, 
			\hat{Q}_{i}^{3z^{2}-r^{2}}, 
			\hat{Q}_{i}^{xy}, 
			\hat{Q}_{i}^{yz},
			\hat{Q}_{i}^{zx} )$, 
is generally applied, which are evaluated as 
	\begin{align}
		\label{eq:quadrupolar_dvector}
		\boldsymbol{Q}_{i}
	= \Braket{\psi_{i}|\hat{\boldsymbol{Q}}_{i}|\psi_{i}}
	= \begin{pmatrix}
			- \left( \left| d_{i, x} \right|^{2} -  \left| d_{i, y} \right|^{2} \right) \\
			- \dfrac{1}{\sqrt{3}} \left( 2 \left| d_{i, z} \right|^{2} - \left| d_{i, x} \right|^{2} - \left| d_{i, y} \right|^{2} \right) \\
			- \left( d_{i, x}^{*} d_{i, y} + d_{i, y}^{*} d_{i, x} \right) \\
			- \left( d_{i, y}^{*} d_{i, z} + d_{i, z}^{*} d_{i, y} \right) \\
			- \left( d_{i, z}^{*} d_{i, x} + d_{i, x}^{*} d_{i, z} \right) \\
		\end{pmatrix}.
	\end{align}
Our simulation is performed on the lattice of $N = L \times L$, 
with $L=12 - 36$ and $L = 8-32$ for the triangular and square lattices, respectively, 
under the periodic boundary condition. 
We combine the conventional Metropolis method with single spin rotations 
and replica exchange method~\cite{Hukushima1996_JPhysSocJpn_65_1604}, taking averages 
over the independent initial configurations using the jackknife resampling, 
each run consisting of $10^{6}$--$10^{7}$ MC steps for thermalization and measurements, respectively.
\par
The sSU(3)-MC considers only the leading terms of the cumulant expansion of the partition function
~\cite{Stoudenmire2009_PhysRevB_79_214436},
which nevertheless incorporate enough amount of quantum and thermal fluctuations 
far beyond the simple mean-field calculation~\cite{footnotemf}. 
In fact, the description Eq.~(\ref{wf}) is considered to describe well the nematic ground states of 
spin-1 BLBQ model on a triangular lattice~\cite{Lauchli2006_PhysRevLett_97_087205,Penc2011_IFM_Chap13_SpinNematics}. 
At finite temperature, the stochastic MC averages further allows us to include the spatial fluctuation effect 
intrinsic to the second order phase transition. 
In the same context, the sSU(3)-MC has distinct advantages over 
the simple classical approximation which treats the spin moments as vectors~\cite{Kawamura2007_JPhysSocJpn_76_073704}; 
in our case the vector $\boldsymbol{d}_{i}$ represents the quantum quadrupole or the local quantum entanglement which orders by the BQ interaction, 
whereas the same interaction in classical method favors the collinear spin orderings~\cite{Stoudenmire2009_PhysRevB_79_214436}.
\par
An artifact of adopting the formally classical Hamiltonian Eq.~(\ref{eq:hamiltonian_1body}) 
is the lack of quantum condensation at low temperature; 
the $T^2$-dependent specific heat characterizing the lowest energy excitations of three-colored Schwinger bosons of the spin-1 BLBQ model
~\cite{Lauchli2006_PhysRevLett_97_087205,Bhattacharjee2006_PhysRevB_74_092406,Penc2011_IFM_Chap13_SpinNematics} is not observed~\cite{entropy_calc}. 
However, in the temperature range at our focus, 
such behavior is easily wiped out by the thermal fluctuations and the collective behavior of quadrupoles become dominant, 
where the validity of Eq.~(\ref{eq:hamiltonian_1body}) shall be recovered. 
In fact, the previous QMC calculations show that the specific heat deviates from $T^{2}$ at $T \sim T_{c}/2$~\cite{Kaul2012_PhysRevB_86_104411,Voll2015_PhysRevB_91_165128}.
\par
\begin{figure}[t]
		\centering
		\includegraphics[width=86mm]{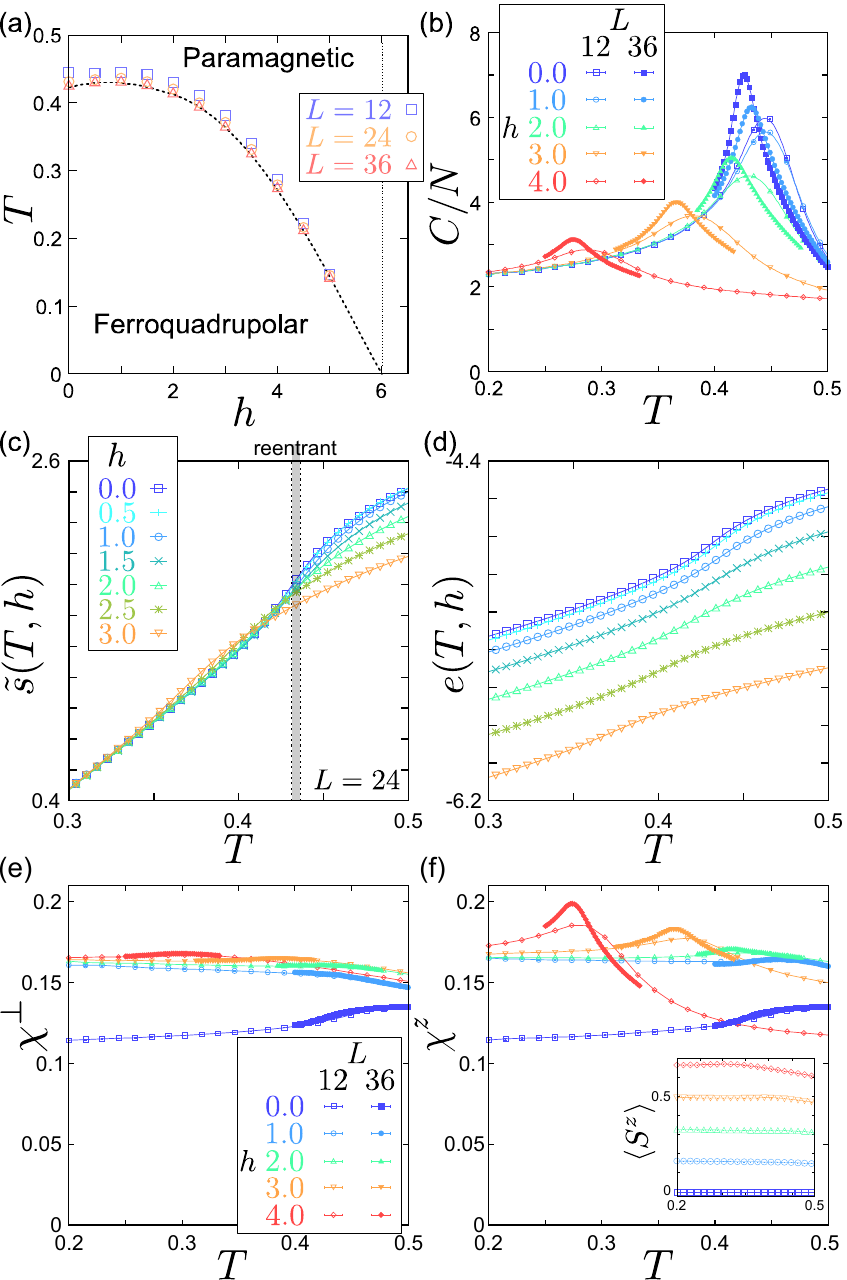}
		\caption{
		Results of the sSU(3)-MC simulations on the triangular lattice at $(J, K) = (0, -1)$. 
	(a) $T$-$h$ Phase diagram.
	Squares, circles, and triangles are the peak positions of the specific heat estimated by the Gaussian kernel approximation in $L = 12, 24, 36$ samples, respectively.
	(b) Temperature dependence of the specific heat $C/N$ for the various values of magnetic field $h$ in $L = 12$ (open) and $36$ (filled) samples.
	(c), (d) Temperature dependence of the entropy $\tilde{s}(T, h)$ and energy $e(T, h)$ per site in $L = 24$ sample.
  (e), (f) Temperature dependence of the magnetic susceptibilities 
	perpendicular and parallel to the magnetic field for the various values of $h$, in $L = 12$ (open) and $36$ (filled) samples.
	The inset in (f) is the temperature dependence of the averaged magentization density for $L = 12$.
		}
		\label{fig:fq_phased}
	\end{figure}

\textit{Ferroquadrupolar phase.} 
We first present the finite-temperature properties of the FQ states. 
Figure~\ref{fig:fq_phased}(a) shows the $T$--$h$ phase diagram for the triangular lattice at $(J, K) = (0, -1)$ 
where the ground state has a FQ ordering. 
We determine the transition temperature $T_{c}$ as the peak position of the specific heat $C/N$; 
Figure~\ref{fig:fq_phased}(b) shows the temperature dependence of $C/N$ in various magnetic fields. 
The peak position first slightly shifts to higher temperature 
and at $h\gtrsim 1$ starts to move rapidly toward lower temperature.  
We compare the results of $L=12$ and 36 to show that the finite size effects are small enough, 
where $T_{c}$ is always slightly lower and the peak-height increases for larger $L$ for all values of $h$ we examined. 
To understand this reentrant behavior of $T_{c}$, we analyze the $T$-dependence of the entropy density $\tilde{s}(T,h)$ 
and the energy density $e(T,h)$ in a weak field as shown in Figs.~\ref{fig:fq_phased}(c) 
and~\ref{fig:fq_phased}(d)~\cite{entropy_calc}; 
$\tilde{s}(T,h)$ does not change much with $h$ for the FQ phase. 
In contrast, the paramagnetic phase rapidly loses its entropy with $h$ since the magnetic moment starts 
to align in the same direction. 
Since $h$ gives a nearly constant shift in the energy $e(T,h)$ common to both phases, 
which is dominated by the Zeeman term, 
the reentrant behavior is understood as an entropic effect. 
\par
\par
Figures~\ref{fig:fq_phased}(e) and (f) show the magnetic susceptibilities 
of the spin component perpendicular and parallel to the magnetic field, 
$\chi^{\perp} = \left( \chi^{x} + \chi^{y} \right) / 2$, and $\chi^{z}$, respectively,
where $\chi^{\alpha} = \beta N \left( \braket{\left(S^{\alpha}\right)^{2}} - \braket{S^{\alpha}}^{2}\right)$.
One finds that $\chi^{z}$ starts to develop a small peak at $T_c$ when $h$ is applied, 
indicating that the finite magnetic moment is induced along the $z$-axis. 
The value of $\chi^{\perp}$ remains almost featureless, but a small structure appears 
at the same position as $\chi^{z}$ for larger $L$. 
The magnetization $\Braket{S^{z}}$ does not depend much on $T$ and its value in the FQ phase is 
in good agreement with the ground state ones, $m = h/\left[ 6 \left( J - K \right) \right] = h/6$ (see inset of Fig.~\ref{fig:fq_phased}(f))~\cite{Lauchli2006_PhysRevLett_97_087205,Penc2011_IFM_Chap13_SpinNematics}.
	\begin{figure}[t]
		\centering
		\includegraphics[width=86mm]{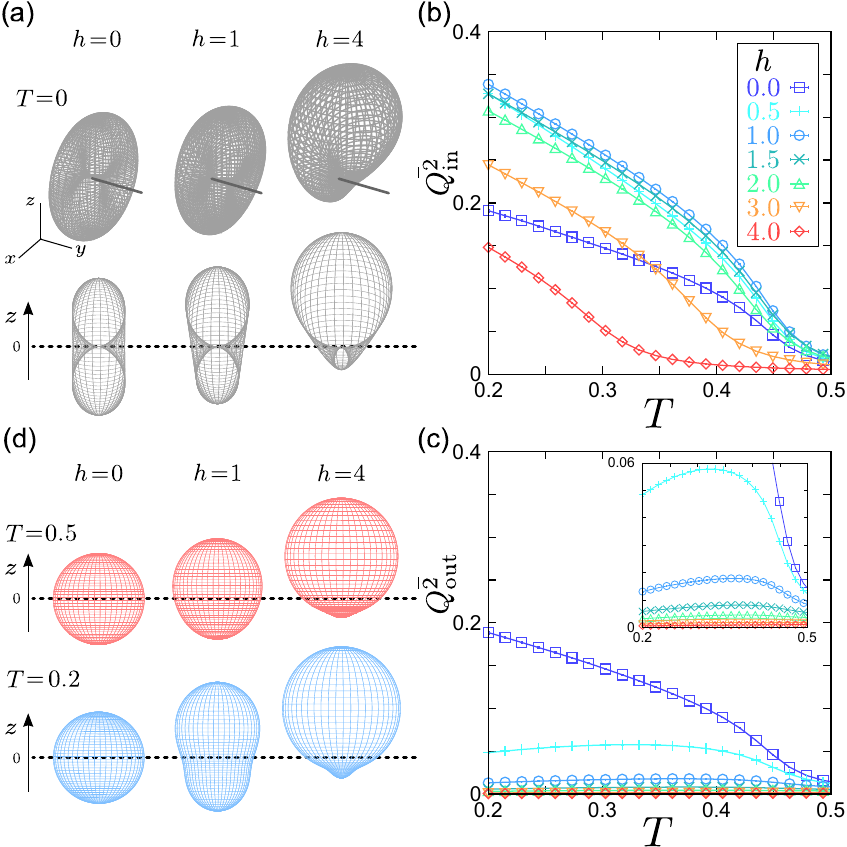}
		\caption{(a) Spin-component distributions
			for the states at $T=0$ in a magnetic field where 
			$\boldsymbol{d} \in {\mathbb{R}}$ (bold line) is pointing in the $y$-direction. 
			The lower panel is the view in the $x$-direction. 
			(b), (c) Temperature dependence of (b) $\bar{Q_{\text{in}}^{2}}$ and 
			(c) $\bar{Q_{\text{out}}^{2}}$ (Eq.~(\ref{eq:q_inout})) in the $L = 12$ triangular lattice.
			The inset in (c) is the enlarged view in the small $\bar{Q_{\text{out}}^{2}}$ region.
			(d) Spin-component distributions
			for the states evaluated by the sSU(3)-MC at $T\neq 0$ and $h=0,1,4$ in $L = 12$ triangular lattice, 
			viewed perpendicular to the $z$-axis. 
			The upper/lower panels belong to the paramagnetic/FQ phases. 
			Since the calculation does not break the O(2) symmetry about the $z$-axis, 
			the quadrupolar moments of $T=0.2$ ones 
			are the averages of those of the symmetry broken ones with 
			its director pointing in a particular direction, 
			which is also consistent with the Mermin-Wagner theorem.
		}
		\label{fig:fq_quad}
	\end{figure}
	\begin{figure}[t]
		\centering
		\includegraphics[width=86mm]{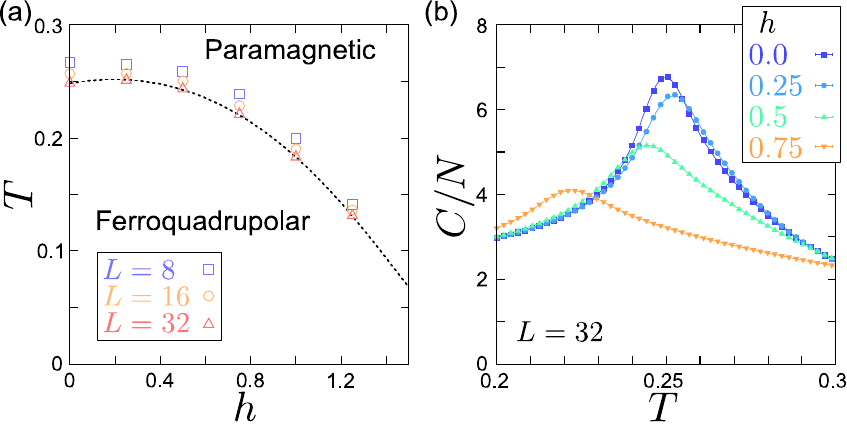}
		\caption{
	(a) $T$-$h$ phase diagram and (b) the $T$-dependence of $C/N$
      for the square lattice at $K/J = 2$ with $J, K < 0$.}
		\label{fig:fq_sq}
	\end{figure}	
\par
We next examine the field-dependence of the quadrupolar moments. 
In the ground state, the magnetic field confines the $\boldsymbol{d}$-vector within the $xy$-space, namely, 
$\Ket{\Psi} = \bigotimes_{i = 1}^{N} \ket{\boldsymbol{d}_{i}}$ with $\ket{\boldsymbol{d}_{i}} = d_{x}\ket{x} + d_{y}\ket{y}$. This is because $h$ couples to $d_{x}^*d_{y}-d_{y}^*d_{x}$  but not with $d_{z}$ 
(see Eq.~(\ref{eq:hamiltonian_1body})). 
When $\boldsymbol{d}$ is real, a pure quadrupole is formed $O(2)$-symmetric about the director $\boldsymbol{d}$, 
as shown in the first column of Fig.~\ref{fig:fq_quad}(a). 
At $h\ne 0$ the emergent imaginary component of $\boldsymbol{d}$ will distort it by shifting its fluctuation center toward the $+z$-direction~\cite{Penc2011_IFM_Chap13_SpinNematics} (see the second and third columns of Fig.~\ref{fig:fq_quad}(a)). 
\par
The relationship between $\boldsymbol{d}$ and the shape of the quadrupole is understood more clearly from 
Eq.~(\ref{eq:quadrupolar_dvector}) as follows; 
when $d_z$ is zero so do $\Braket{Q^{yz}}$ and $\Braket{Q^{zx}}$, 
while $\langle Q^{x^{2}-y^{2}}\rangle$ and $\Braket{Q^{xy}}$ 
which consist only of $x$ and $y$ elements of $\boldsymbol{d}$ can respond to $h$. 
Based on this consideration, we define two kinds of squared quadrupolar moments, 
	\begin{align}
		\label{eq:q_inout}
		\bar{Q_{\text{in}}^{2}}
	& = \dfrac{1}{2}\left( \left(Q^{x^{2}-y^{2}} \right)^{2} + \left( Q^{xy} \right)^{2} \right),
 \notag \\
		\bar{Q_{\text{out}}^{2}}
	& = \dfrac{1}{2} \left( \left( Q^{yz} \right)^{2} + \left( Q^{zx} \right)^{2} \right),
	\end{align}
where we straightforwardly find $\bar{Q_{\text{in}}^{2}} \neq 0$ and $\bar{Q_{\text{out}}^{2}} = 0$ at $T = 0$.
\par
Temperature dependences of $\bar{Q_{\text{in}}^{2}}$ and $\bar{Q_{\text{out}}^{2}}$ are shown 
in Figs.~\ref{fig:fq_quad}(b) and \ref{fig:fq_quad}(c). 
In a weak field, $\bar{Q_{\text{in}}^{2}}$ increases first and then at around $h\gtrsim 1$ starts to decrease with $h$.
Whereas, $\bar{Q_{\text{out}}^{2}}$ takes a small but finite value 
at small $h$ and $T\ne 0$ because of the thermal flucuation, 
and becomes $\bar{Q_{\text{out}}^{2}} = 0$ at $T = 0$. 
Notice that the apparently large values of $\bar{Q_{\text{out}}^{2}}$ at $h=0$ simply because the moments are decoupled to the spatial coordinate. 
Once the field increases to $h \gtrsim 1$, $\bar{Q_{\text{out}}^{2}}$ is suppressed to zero for all temperatures. 
These results indicate that there is a crossover at $h \sim 1$ 
from the low-field regime with a robust quadrupolar moment 
to the high field regime where the quadurupolar moment is rapidly suppressed with field.
It apparently links with the reentrant behavior of $T_{c}$ at $h \lesssim 1$.
\begin{figure}[t]
	\centering
	\includegraphics[width=86mm]{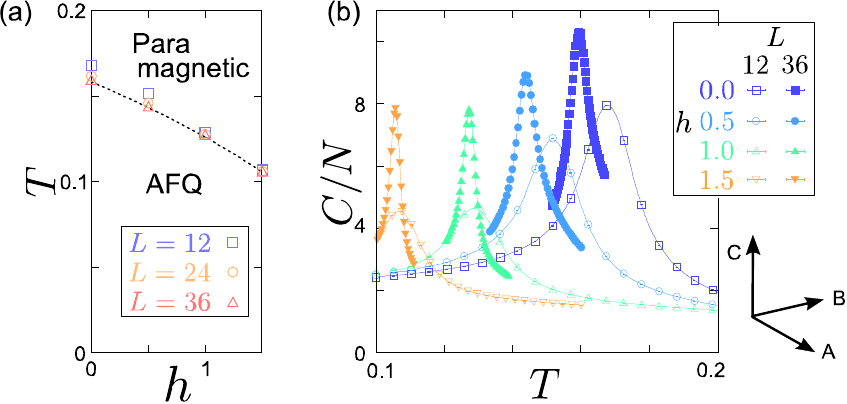}
	\caption{
	(a) $T$-$h$ phase diagram and (b) the $T$-dependence of $C/N$ 
      for the triangular lattice with $L = 12$ (open) and $36$ (filled). 
      The antiferroquadrupolar phase at $K/J = 2$ with $J, K > 0$ is studied~\cite{afq_parameter}, 
      where A and B-sublattices gradually develop a magnetic moment similarly to that of the aforementioned FQ phase at $T=0$. 
      In the higher field range of $h=3/\sqrt{5}\approx 1.34$ to $3(1-1/\sqrt{5})\approx 1.658$ a 2/3 plateau phase appears, 
where the $\ket{\boldsymbol{d}_{i_{\text{C}}}} = \ket{z}$ remains and the spins on A and B sublattices are fully polarized. 
			}
			\label{fig:afq}
		\end{figure}
Such particular field-dependence of the quadrupolar moment is vidualized in Fig.~\ref{fig:fq_quad}(d). 
Compared to the ones at $T=0$, the quadrupoles at $h=1$ and $T=0.2$ has a gourd-shape 
which indicates that $\boldsymbol{d}$ cants slightly off the $xy$-plane to maximize the fluctuation. 
\par
\textit{Remarks.}
The geometry of the lattice does not seem to play any intrinsic role. 
As shown in Figs.~\ref{fig:fq_sq}(a) and \ref{fig:fq_sq}(b), 
we also perform the similar calculation for the square lattice at $K/J = 2$ with $J, K < 0$~\cite{fq_square_parameter}, 
finding a reentrant behavior at $h \lesssim 0.25$~\cite{square_supplement}.
\par
In both Eqs.~(\ref{ham}) and~(\ref{eq:hamiltonian_1body}), the continuous symmetry exists, 
in which case a bulk finite-temperature phase transition is prohibited by 
the Mermin--Wagner theorem~\cite{Mermin1966_PhysRevLett_17_1307}. 
However, the specific heat shows a peak at $T_{c}$ not only in our case but in the previous QMC calculations, 
which indicates the development of an intrinsic quasi-long range ordering~\cite{KT_transition}. 
The growth of correlation length at $T_{c}$, even if it does not diverge, 
will drive the system to a long range ordering once we stack the two-dimensional layer and 
include the inter-layer coupling, which is the situation expected for the actual material systems. 
\par
\textit{Antiferroquadrupolar phase.} 
Finally, we briefly discuss the finite temeprature properties and its field dependence for the antiferroquadrupolar (AFQ) phase. 
In the ground state of the triangular lattice, the AFQ is realized 
at $0 < J < K$~\cite{Tsunetsugu2006_JPhysSocJpn_75_083701,Lauchli2006_PhysRevLett_97_087205,Tsunetsugu2007_JPhysCondensMatter_19_145248}. 
The quadrupolar moments form a three-sublattice structure described by 
$\ket{\Psi} = \bigotimes_{i = 1}^{N/3} \left( \ket{\boldsymbol{d}_{i_{\text{A}}}} \otimes\ket{\boldsymbol{d}_{i_{ \text{B}}}}\otimes\ket{\boldsymbol{d}_{i_{\text{C}}}} \right)$,
where $i_{\gamma}$ ($\gamma = \text{A, B, C}$) denotes the lattice sites with sublattice indices. 
At $h\ne 0$ and $T = 0$, the spatial direction of $\boldsymbol{d}_{i_{\gamma}}$ is fixed,  
and two of three components, $\ket{\boldsymbol{d}_{i_{\text{A}}}} = \ket{x}$ and $\ket{\boldsymbol{d}_{i_{\text{B}}}} = \ket{y}$, 
behave similarly to the $\boldsymbol{d}$-vector of the FQ phase in a magnetic field, i.e., $d_{x}$ and $d_{y}$ become complex numbers, 
while $\ket{\boldsymbol{d}_{i_{\text{C}}}} = \ket{z}$ remains real~\cite{Lauchli2006_PhysRevLett_97_087205}. 
Figure~\ref{fig:afq}(a) shows the $T$--$h$ phase diagram; 
$T_c$ decreases with increasing $h$, which can be detected clearly by the temperature dependence of the specific heat in 
Fig.~\ref{fig:afq}(b) for the different values of $h$ in $L = 12$ and $36$ samples. 
Although the quadrupoles on the A- and B-sublattices behave similar to that of the FQ phase, 
the difference lies in that they do not acquire a finite $d_{z}$ and continue pointing within the $xy$-plane. 
In fact, the slight increase in $T_c$ at small $h$ is not observed in the case of AFQ. 
This is because the correlation between the quadrupoles of different sublattices do not allow the directors 
to cant off the original angle. The reentrant $T_c$ is thus not observed for translational symmetry breaking AFQ ordering. 
\par
\textit{Conclusion.}
We disclosed the slight reentrant behavior of the paramagnetic-to-ferroquadrupolar spin nematic transition temperature 
in a weak magnetic field. 
Although both phases develop almost the same amount of magnetic moment by the field, 
only the nematic phase remains robust by modifying the shape of the quadrupolar moment toward the in-plane direction, 
and the counterpart paramagnet rapidly loses its entropy. 
One finds a common feature in $^3$He in that the entropies of solid and liquid phases 
comes from the weakly interacting spin-1/2 nuclei and the condensed liquid of spins, respectively, 
although in their case the two phases are separated by the first order transition. 
If we regard the enthalpy $+PV$ as our Zeeman term, $-\braket{S^{z}}h$, 
the Pomeranchuk effect, namely $^3$He is cooled by tracing the solid-to-liquid transition temperature 
in applying the pressure, would explain our increase of $T_c$ with $h$. 
In both cases, the decrease of relative entropy per unit (volume or spin) of solid to liquid plays a crucial role to raise $T_c$. 
Similar field reentrant phase diagram is found in the uud plateau phase of the triangular lattice quantum magnet, 
where a large quantum fluctuation supported by the geometrical frustration shall play a similar role to present quadrupolar fluctuation~\cite{Tsujii2007_PhysRevB_76_060406}. 
We thus conclude that the present phenomenon not only serves as a clue to explore elusive nematic phase, 
but adds to the list of universal entropic effect intrinsic to the physics of quantum liquid versus paramagnet. 
\par
\textit{Acknowledgements.}
We thank Hikomitsu Kikuchi, Yutaka Fujii, Karlo Penc, Katsuhiro Morita, and Tsutomu Momoi for fruitful discussions and comments. 
This work is supported by JSPS KAKENHI Grants No. JP17K05533, No. JP18H01173,
No. JP17K05497, and No. JP17H02916.
Part of the calculations is done using the facilities of the Supercomputer Center, Institute for Solid State Physics, University of Tokyo.

\begin{thebibliography}{49}
\bibitem [{foo()}]{footnotemf}%
  \BibitemOpen
  \href@noop {} {}\bibinfo {note} {The simple mean-field approach assumes that
  $\boldsymbol{d}_{i}$ in Eq.~(\ref{wf}) is spatially uniform, whereas
  sSU(3)-MC allows $\boldsymbol{d}_{i}$ to vary both in space and in Monte
  Carlo timescale, which takes account of the fluctuation other than and the
  quantum condensation effect referred in~\cite{entropy_calc}. There, the
  fluctuation originating from the inter-site quantum entanglement can be
  replaced by the Monte Carlo averages. Particularly for the class of trivial
  long range order which are well approximated by the product state of the
  locally entangled unit (e.g., product of quadrupoles in Eq.~(\ref{wf})), the
  quantum and thermal fluctuation cannot be discriminated, and the present
  treatment shall give a good description.}\BibitemShut {Stop}%
\end{thebibliography}
%

\clearpage
\clearpage
\pagestyle{empty}
\setlength{\textwidth}{1.15\textwidth}
\begin{figure*}
	\vspace{-20mm}\hspace{-30mm}
	\centering
	\includegraphics[page=1]{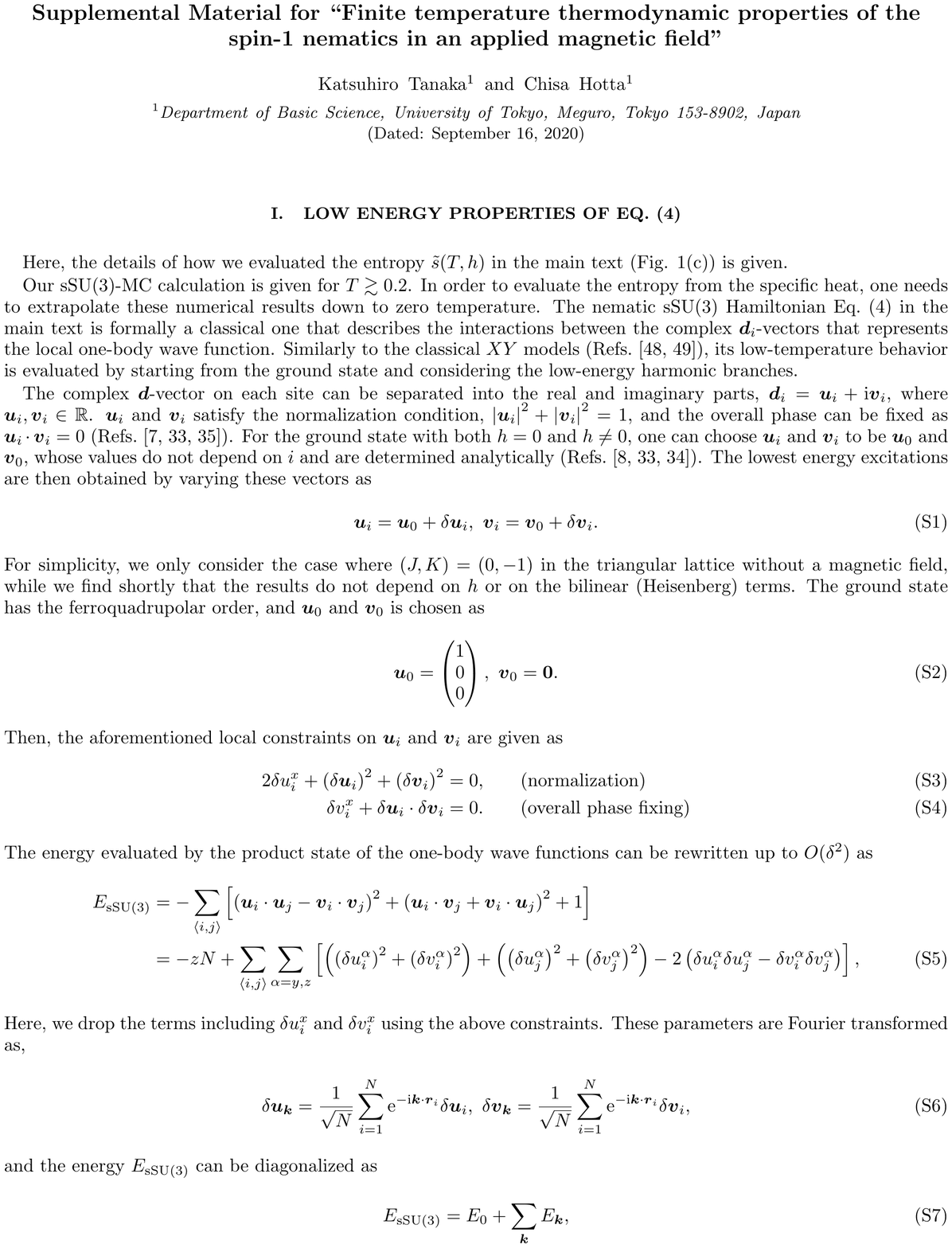}
\end{figure*}
\begin{figure*}
	\vspace{-15mm}\hspace{-30mm}
	\centering
	\includegraphics[page=2]{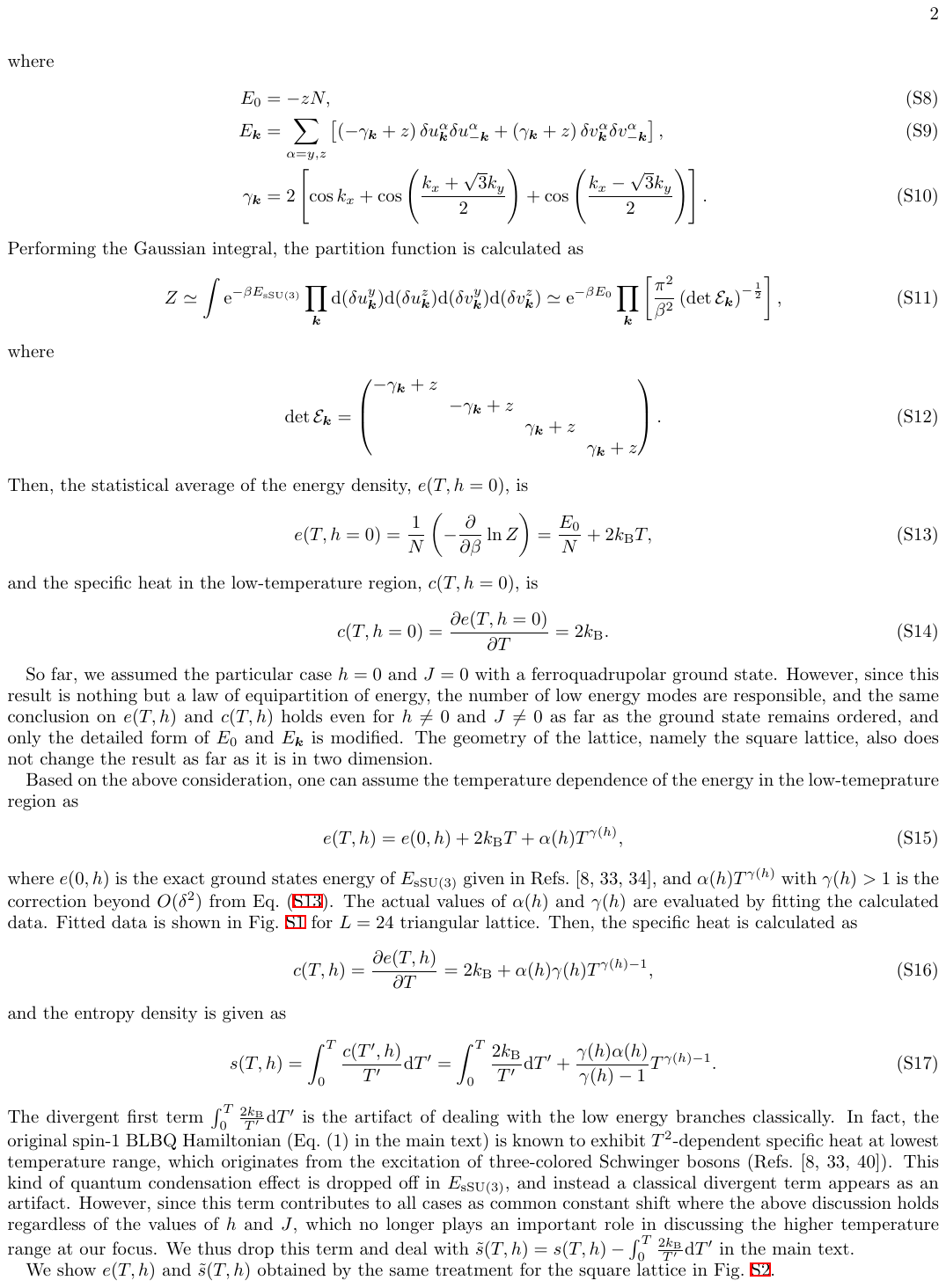}
\end{figure*}
\begin{figure*}
	\vspace{-15mm}\hspace{-30mm}
	\centering
	\includegraphics[page=3]{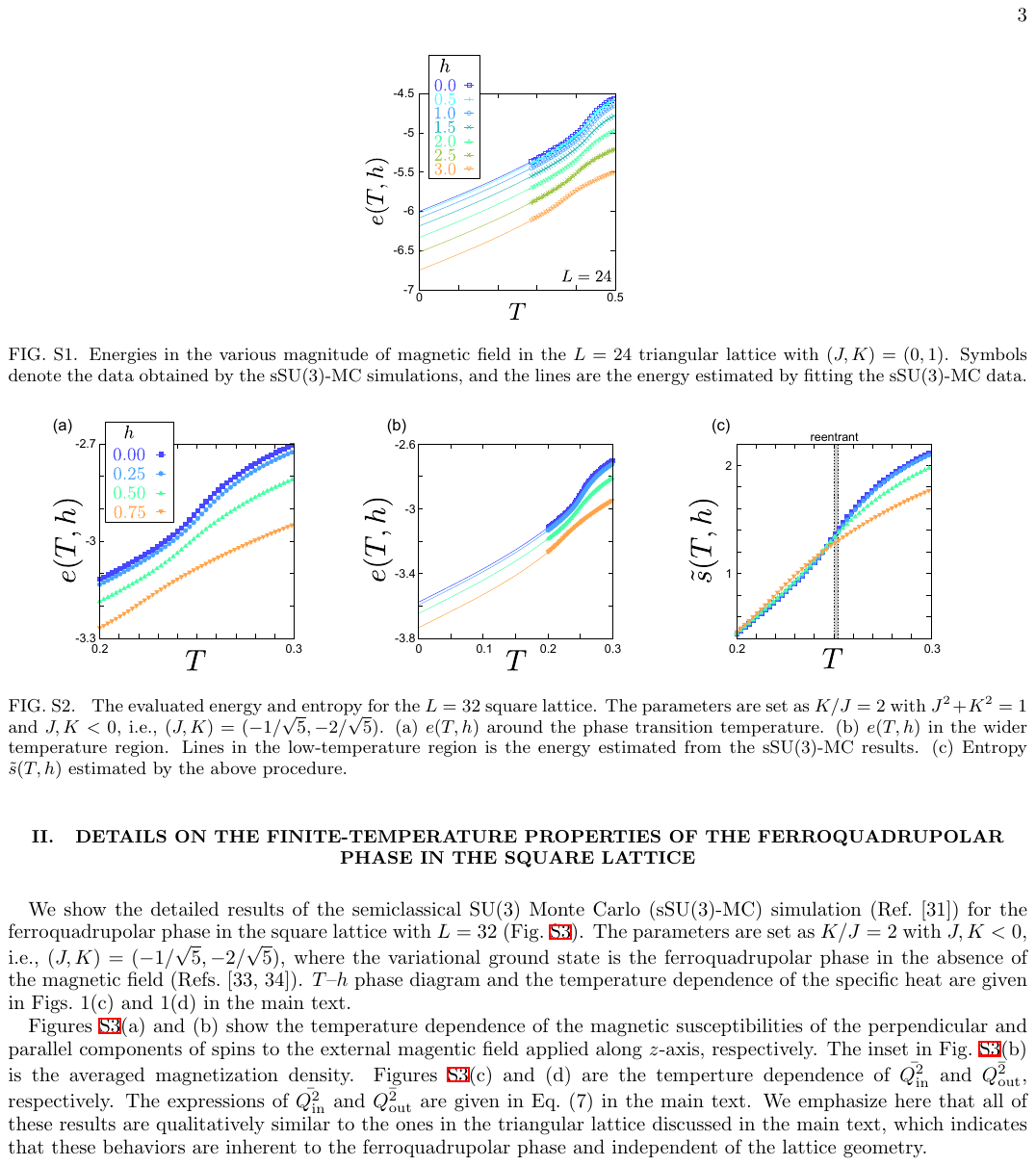}
\end{figure*}
\begin{figure*}
	\vspace{-15mm}\hspace{-30mm}
	\centering
	\includegraphics[page=4]{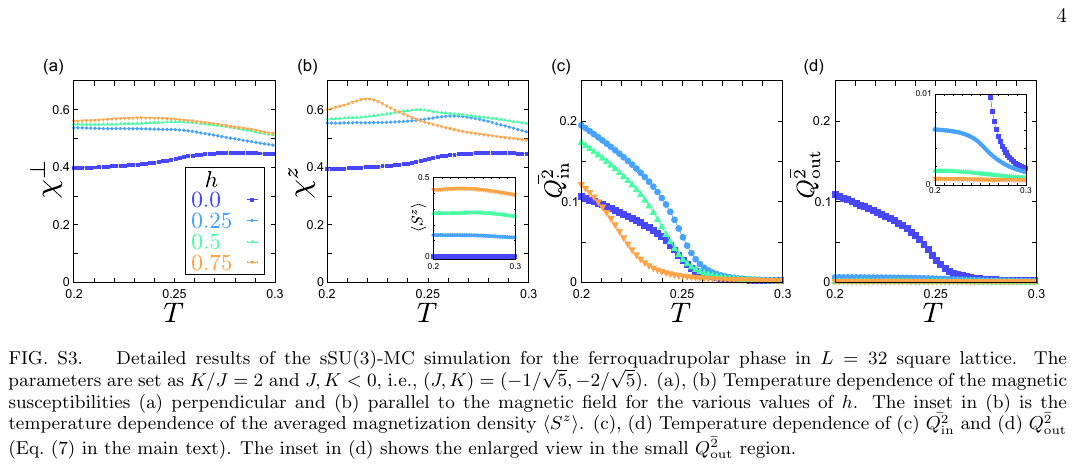}
\end{figure*}
\end{document}